\documentclass[journal=nalefd,manuscript=article,layout=twocolumn]{achemso}

\usepackage[version=4]{mhchem} 
\usepackage{graphicx}          
\usepackage{amsmath,amssymb}   
\usepackage{siunitx}           
\usepackage{hyperref}          

\title{Helicity-Controlled Hall Transport\\ in Hybrid Topological Magnetic Textures}

\author{Anton~V.~Hlushchenko}
\email{glushchenko.ant@gmail.com}
\affiliation{National Science Center "Kharkiv Institute of Physics and Technology",\ \\
61108 Kharkiv, Ukraine}

\author{Julia Kharlan}
\affiliation{ISQI, Faculty of Physics, Adam Mickiewicz University,\
61-614 Pozna{\'n}, Poland}
\alsoaffiliation{V.G. Baryakhtar Institute of Magnetism of the NAS of Ukraine,\
03142 Kyiv, Ukraine}
\alsoaffiliation{G. V. Kurdyumov Institute for Metal Physics of the NAS of Ukraine,\
02000 Kyiv, Ukraine}

\author{Mykhailo~I.~Bratchenko}
\affiliation{National Science Center "Kharkiv Institute of Physics and Technology",\ \\
61108 Kharkiv, Ukraine}

\author{Aleksei~V.~Chechkin}
\affiliation{National Science Center "Kharkiv Institute of Physics and Technology",\ \\
61108 Kharkiv, Ukraine}
\alsoaffiliation{Max Planck Institute of Microstructure Physics,\
06120 Halle, Germany}
\alsoaffiliation{Faculty of Pure and Applied Mathematics,\ \\
Wroc{\l}aw University of Science and Technology, 50-370 Wroc{\l}aw, Poland}

\author{Jaros{\l}aw W. K{\l}os}
\affiliation{ISQI, Faculty of Physics, Adam Mickiewicz University,\
61-614 Pozna{\'n}, Poland}

\author{Oleg~A.~Tretiakov}
\email{o.tretiakov@unsw.edu.au}
\affiliation{School of Physics, The University of New South Wales,\
2052 Sydney, Australia}

\usepackage{bm}
\usepackage{amsthm}
\usepackage{multirow}
\usepackage{subfigure}
\usepackage{float}
\usepackage{xcolor}
\usepackage[normalem]{ulem}
\usepackage{fancyhdr}

\begin{document}

\begin{abstract}
Topological magnetic textures can serve as information carriers driven by spin currents. However, their motion is generally accompanied by a Hall effect that deflects them from the current direction, limiting transport efficiency and controllability.
We develop a unified spin-space transformation framework enabling a systematic study of hybrid spin textures with different helicities, such as skyrmions, antiskyrmions, bimerons, and antibimerons.
Combining analytical theory with micromagnetic simulations, we establish the relation between helicity and current-driven transport. An analytical solution of the generalized Thiele equation identifies helicity as a geometric control parameter and yields a simple expression for the Hall angle, enabling its continuous tuning, complete suppression, and deterministic steering along arbitrary in-plane directions. Micromagnetic simulations confirm that the generated spin textures remain stable under Landau--Lifshitz--Gilbert dynamics and validate the analytical predictions. These results establish helicity as a versatile control parameter for programmable transport of topological magnetic textures.

\end{abstract}








%

\section{Introduction}
The rapid development of skyrmionics has demonstrated that magnetic skyrmions represent only one member of a much broader family of topological magnetic textures. Depending on the crystal symmetry, magnetic interactions, and external stimuli, a variety of topological spin textures have been theoretically predicted and experimentally realized, including Bloch- and N\'eel-type skyrmions \cite{Yu_Nature_2010,Heinze_NatPhys_2011,Seki_Science_2012}, antiskyrmions \cite{Nayak2017}, bimerons \cite{Tretiakov_PhysRevB_2019,Bhukta_NatCommun_2024,Castro_NanoLett_2025,Vorobyev2026}, synthetic antiferromagnetic skyrmions \cite{FernandezPacheco_NatMater_2019,Pham_Science_2024}, and three-dimensional topological textures such as hopfions \cite{Kent2021}. Beyond the conventional Dzyaloshinskii--Moriya interaction (DMI), skyrmions can also be stabilized by Ruderman--Kittel--Kasuya--Yosida (RKKY) interactions \cite{Soumyanarayanan_Science_2018}, dipolar interactions \cite{Moutafis_PRB_2017}, and external magnetic fields in hybrid magnetic--superconducting nanostructures \cite{79d8-gmcs}. These diverse  mechanisms for the formation of metastable topological textures considerably expand the design space of topological spintronic devices \cite{Zhang2020, Gobel2021}.

One of the major challenges for current-driven skyrmion-based devices is the skyrmion Hall effect, which leads to undesirable transverse motion and may result in skyrmion annihilation at racetrack edges \cite{Litzius2017, Jiang2017,Yang_APR_2024}. Consequently, considerable efforts have been devoted to controlling or suppressing the Hall deflection through various material and device engineering strategies. These include tailoring the spin polarization generated by spin--orbit torques via crystal-symmetry engineering and anisotropic spin Hall conductivity tensors \cite{Seemann_PRB_2015,Wimmer_PRB_2015,Zhang_SciAdv_2016,Zhang_PRB_2017,Goebel_PRB_2019}, engineering the spin--orbit interaction itself to modify the Hall response \cite{Akosa_PRApp_2019}, and employing antiferromagnetic \cite{Barker2016, Bessarab_PhysRevB_2019} or synthetic antiferromagnetic structures that compensate the gyrotropic response of coupled skyrmions while simultaneously enabling high-speed current-driven transport \cite{Akosa_PRL_2018,Pham_Science_2024,deSouzaSilva_PRL_2025}.

More recently, helicity has emerged as an additional degree of freedom
for controlling the transport properties of magnetic textures,
providing a route toward engineering the skyrmion Hall effect and
current-driven dynamics
\cite{Diaz_JPCM_2016,Wu_PRB_2017,Kim_PRB_2018_HybridDMI,Li_AdvMater_2019,Liu_APL_2023,Akhir_JPhysD_2024,Chang2025}. At the same time, rapid experimental and theoretical progress has considerably expanded the family of accessible topological magnetic textures beyond conventional skyrmions to include bimerons, antiskyrmions, hybrid skyrmions, and related topological spin configurations
\cite{Legrand_SciAdv_2018,Carvalho_PRMater_2021,Peng_NatNano_2020,Ohara2022,Amin2023,Li_2026,Chang2025}. These developments have demonstrated that different magnetic textures can often be transformed into one another through modifications of the underlying magnetic interactions or external control parameters such as magnetic fields, strain, and electric fields
\cite{Goebel_PRB_2019,Goerzen2026,Jena2026,Fe2026}. Collectively, they therefore provide a natural basis for a unified spin-space framework capable of describing the entire family of
topological magnetic textures within a common theoretical formalism.

In this paper we employ orthogonal $O(3)$ spin-space transformations to systematically
generate families of skyrmions, bimerons, antiskyrmions, and antibimerons
from a reference N'eel skyrmion. The proposed transformation is accompanied by a corresponding covariant transformation of the micromagnetic Hamiltonian, specifically of the magnetic-anisotropy and DMI energy-density terms, thereby preserving the metastability of the transformed spin textures. Furthermore, this framework enables continuous interpolation between the conventional N'eel and Bloch configurations, allowing topological textures with arbitrary helicity to be generated and providing a generalized micromagnetic description of hybrid topological magnetic textures.

Within this framework, we investigate both analytically and numerically the current-driven dynamics of the generated spin textures to establish the relation between helicity and their transport properties, with particular emphasis on the control of Hall transport. We show that the generalized Thiele equation yields an analytical expression for the Hall angle predicting that the texture propagation angle varies linearly with helicity, with a slope set by the topological charge, whereas the speed is independent of helicity at fixed micromagnetic parameters. Helicity can therefore eliminate the transverse velocity or steer a texture along any in-plane direction under a fixed current. Micromagnetic simulations based on Landau--Lifshitz--Gilbert (LLG) dynamics quantitatively reproduce the predicted longitudinal and transverse velocities over the full helicity range. Our results establish a general strategy for controlling the Hall transport of topological magnetic textures through helicity engineering.

\begin{figure*}
\centering
\includegraphics[width=1.0\linewidth]{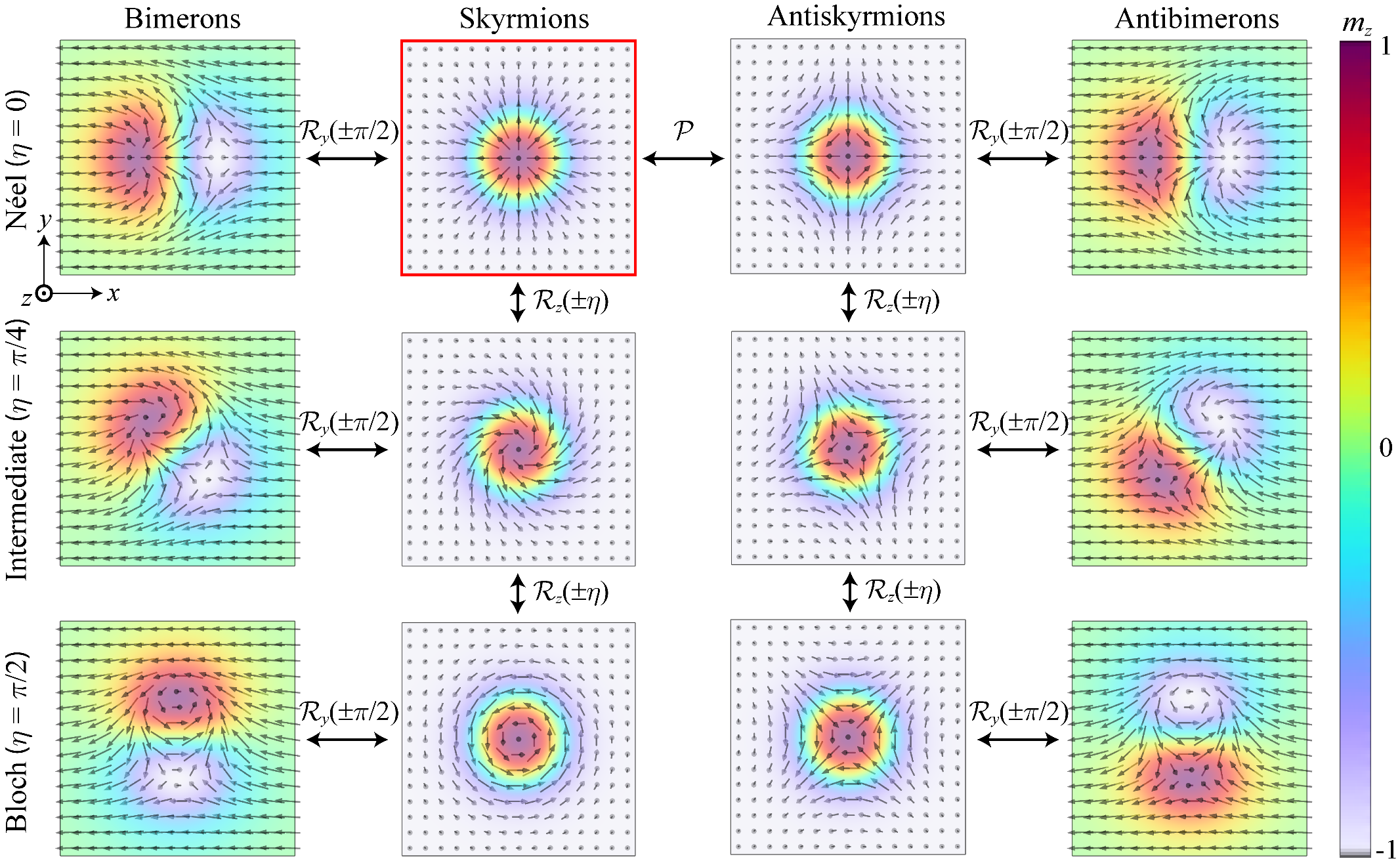}
\caption{\label{fig:spin_transformations}
Schematic illustration of the proposed spin-space transformation
framework for generating the complete family of hybrid topological
magnetic textures. The red frame highlights the reference N\'eel
skyrmion, which serves as the starting point for the transformations.
The four columns present the families of bimerons, skyrmions,
antiskyrmions, and antibimerons, while the three rows correspond to the
N\'eel ($\eta=0$), intermediate ($\eta=\pi/4$), and Bloch
($\eta=\pi/2$) configurations. The vertical arrows denote the helicity
rotation $\mathcal{R}_z(\pm\eta)$, which continuously generates hybrid magnetic
textures with arbitrary helicity within each family. The horizontal
arrows denote the spin-space rotation $\mathcal{R}_y(\pm\pi/2)$, which maps the
skyrmion and antiskyrmion families onto the corresponding bimeron and
antibimeron families, respectively, whereas the transformation operator
$\mathcal P$ reverses the winding direction and the topological charge,
mapping skyrmions onto antiskyrmions and bimerons onto antibimerons.
}
\end{figure*}

\section{Unified framework for topological magnetic textures}
\label{sec:unified_framework}

Here we introduce the proposed spin-space transformation formalism,
which provides a unified framework for generating different classes of
topological magnetic textures from a single reference configuration.
The complete transformation scheme and the resulting family of magnetic
textures are illustrated in Fig.~\ref{fig:spin_transformations}.
Throughout this work, the canonical N\'eel skyrmion
\cite{Sampaio2013,Nagaosa2013,Rohart2013,Zhang2020,LU2023171044} is chosen as the reference texture,
while all remaining textures are generated through successive elementary
spin-space operations. The derivation of the reference ansatz and the
corresponding generalized DMI energy densities is provided in the
Supplementary Information, Secs.~S1--S3. Here, we summarize the
transformation rules and the resulting micromagnetic energy functional.

The normalized magnetization is represented by the vector field
\begin{equation}
\mathbf m(r,\varphi)
=
\left(
m_x(r,\varphi),
m_y(r,\varphi),
m_z(r,\varphi)
\right)^{\rm T},
\end{equation}
where $(r,\varphi)$ denote the polar coordinates in the film plane,
with the origin located at the center of the topological texture.

The micromagnetic energy is written as
\begin{equation}
E[\mathbf m]
=
\int_V
\left(
w_{\rm ex}
+
w_{\rm ani}
+
w_{\rm DMI}
\right)dV,
\end{equation}
where $w_{\rm ex}$, $w_{\rm ani}$, and $w_{\rm DMI}$ denote the
exchange, anisotropy, and DMI energy densities, respectively.
The exchange energy density is
$w_{\rm ex}=A(\nabla\mathbf m)^2$ and is invariant under the
global spin-space transformations considered here. The anisotropy
energy retains the same functional form,
$w_{\rm ani}=-K_u(\mathbf m\cdot\mathbf n)^2$,
with the easy-axis direction $\mathbf n$ transformed consistently with
the magnetization. In contrast, the DMI contribution
$w_{\rm DMI}$ is transformed into a generalized energy density whose
form depends on the corresponding magnetic texture. The explicit
expressions for the generalized DMI energy densities are provided in
the Supplementary Information, Secs.~S1--S3.
We do not explicitly account for the full dipolar interaction, except
for its contribution to the shape anisotropy of the ferromagnetic film,
which can be incorporated into the effective uniaxial anisotropy.
Nonlocal dipolar interactions, finite-boundary effects, and
material-specific anisotropies may deform the transformed textures and
introduce quantitative deviations from the ideal covariance relations.

The reference N\'eel skyrmion adopted throughout this work is based on
the analytical model of Refs.~\cite{Rohart2013,LU2023171044}, which
provides an approximate equilibrium profile for an isolated skyrmion.
The proposed transformations are applied simultaneously to the
magnetization field and to the corresponding interaction terms entering
the energy functional. For an exact stationary solution, this
covariant transformation maps the solution of the reference
micromagnetic model onto a stationary solution of the corresponding
transformed model. In the present work, the analytical reference
profile is used to construct the transformed configurations, which are
subsequently relaxed numerically in the micromagnetic simulations.
Below we summarize the resulting micromagnetic models for skyrmions,
bimerons, antiskyrmions, and antibimerons, establishing a unified
description of the four texture families.

\begin{table*}[t]
\centering
\caption{\label{tab:transformation_rules}
Summary of the spin-space transformations and the corresponding
generalized DMI energy densities for the four families of topological
magnetic textures.
}
\renewcommand{\arraystretch}{1.35}

\begin{tabular}{ccc}
\hline
Texture
&
Spin-space transformation
&
$w_{\rm DMI}(\eta)/D_{\rm DMI}$
\\
\hline

Skyrmion
&
$\mathcal{R}_z(\eta)$
&
$\displaystyle
\left(-L_{zx}^{(x)}+L_{yz}^{(y)}\right)\cos\eta
+
\left(L_{yz}^{(x)}+L_{zx}^{(y)}\right)\sin\eta
$
\\[4mm]

Bimeron
&
$\mathcal{R}_y(\pi/2)\mathcal{R}_z(\eta)$
&
$\displaystyle
\left(-L_{zx}^{(x)}-L_{xy}^{(y)}\right)\cos\eta
+
\left(-L_{xy}^{(x)}+L_{zx}^{(y)}\right)\sin\eta
$
\\[4mm]

Antiskyrmion
&
$ \mathcal{R}_z(\eta)\mathcal{P}$
&
$\displaystyle
\left(-L_{zx}^{(x)}-L_{yz}^{(y)}\right)\cos\eta
+
\left(L_{yz}^{(x)}-L_{zx}^{(y)}\right)\sin\eta
$
\\[4mm]

Antibimeron
&
$ \mathcal{R}_y(\pi/2)\mathcal{R}_z(\eta)\mathcal{P}$
&
$\displaystyle
\left(-L_{zx}^{(x)}+L_{xy}^{(y)}\right)\cos\eta
+
\left(-L_{xy}^{(x)}-L_{zx}^{(y)}\right)\sin\eta
$
\\
\hline

\end{tabular}
\end{table*}

\textit{Skyrmions.}
The construction starts from the reference N\'eel skyrmion introduced in
Supplementary Information S1.  We consider orthogonal
spin-space transformations $O\in O(3)$, with $\det O=+1$ and $\det
O=-1$ corresponding to proper and improper transformations,
respectively.
The complete skyrmion family is generated by the proper rotation about
the $z$ axis,
\begin{equation}
\mathbf m_{\rm Sky}(r,\varphi,\eta)
=
\mathcal{R}_z(\eta)\,
\mathbf m_{\rm SkN}(r,\varphi),
\label{eq:Rz_sk}
\end{equation}
where $\eta$ is the helicity angle. Applying this transformation gives
\begin{equation}
\mathbf m_{\rm Sky}(r,\varphi,\eta)
=
\begin{pmatrix}
\sin\theta(r)\cos(\varphi+\eta)\\
\sin\theta(r)\sin(\varphi+\eta)\\
\cos\theta(r)
\end{pmatrix},
\label{eq:ansatz_skyrmion}
\end{equation}
while the radial profile $\theta(r)$ remains unchanged. The helicity
spans $\eta\in[0,2\pi)$, with the N\'eel- and Bloch-type configurations
corresponding to $\eta=0,\pi$ and $\eta=\pi/2,3\pi/2$, respectively.

The exchange interaction is invariant under this global spin-space
rotation, while the perpendicular easy axis
$\mathbf n=\hat{\mathbf z}$ remains unchanged. Consequently, the
rotation modifies only the form of the DMI. For the reference N\'eel and Bloch contributions \cite{TopologyInMagnetism2018,Niu_NatCommun_2024}, we use
\begin{align}
w_{\rm SkN}
&=
D_{\rm DMI}
\Big[
-m_z\partial_xm_x
-
m_z\partial_ym_y
\nonumber\\
&\qquad
+
m_x\partial_xm_z
+
m_y\partial_ym_z
\Big]
\nonumber\\
&=
D_{\rm DMI}
\left[
-L_{zx}^{(x)}
+
L_{yz}^{(y)}
\right],
\\[1ex]
w_{\rm SkB}
&=
D_{\rm DMI}
\Big[
-m_z\partial_xm_y
+
m_z\partial_ym_x
\nonumber\\
&\qquad
-
m_x\partial_ym_z
+
m_y\partial_xm_z
\Big]
\nonumber\\
&=
D_{\rm DMI}
\left[
L_{yz}^{(x)}
+
L_{zx}^{(y)}
\right],
\end{align}
where $L_{ij}^{(k)}=m_i\partial_km_j-m_j\partial_km_i$
are the Lifshitz invariants~\cite{Lifshitz1941I,Lifshitz1941II,Yang2023,Niu_NatCommun_2024}.
The generalized DMI energy density is then
\begin{equation}
w_{\rm DMI}^{\rm Sky}(\eta)
=
w_{\rm SkN}\cos\eta
+
w_{\rm SkB}\sin\eta.
\label{eq:DMI_sk}
\end{equation}
Thus, the helicity rotation simultaneously transforms the magnetization
and the corresponding DMI, continuously connecting the N\'eel, Bloch,
and intermediate hybrid skyrmions. The complete derivation of the
generalized DMI is given in Supplementary Information S2. Equation~\ref{eq:DMI_sk}
gives the generalized DMI energy density for the skyrmion family. For a
magnetic film of thickness $\Delta_t$, the interfacial and bulk DMI
contributions are characterized by the effective coefficients
$D_s/\Delta_t$ and $D_v$, respectively. Their combined magnitude defines
the DMI energy scale $D_{\rm DMI}$, while their relative weights determine
the helicity
$\eta=\operatorname{atan2}(D_v\Delta_t,D_s)$.

\textit{Bimerons.}
The bimeron family is obtained from the skyrmion family by the fixed
spin-space rotation
\begin{equation}
\mathbf m_{\rm Bim}(r,\varphi,\eta)
=
\mathcal{R}_y\!\left(\frac{\pi}{2}\right)
\mathbf m_{\rm Sky}(r,\varphi,\eta).
\label{eq:Ry_bimeron}
\end{equation}
This rotation maps the perpendicular easy axis
$\mathbf n=\hat{\mathbf z}$ onto the in-plane direction
$\mathbf n=\hat{\mathbf x}$ while preserving the form of the exchange
and anisotropy energies. The DMI, in contrast, is transformed together
with the spin configuration and acquires the corresponding bimeron
form. In the DMI-tensor representation, this transformation is
implemented by applying the same spin-space rotation to the DMI
coefficients. The resulting generalized DMI is summarized in
Table~\ref{tab:transformation_rules}; its explicit derivation is given
in Supplementary Information S2.

\textit{Antiskyrmions.}
Antiskyrmions are generated from the reference skyrmion by the improper
spin-space transformation
$\mathcal{P}:(m_x,m_y,m_z)\rightarrow(m_x,-m_y,m_z)$, followed by the helicity
rotation,
\begin{equation}
\mathbf m_{\rm ASky}(r,\varphi,\eta)
=
\mathcal{R}_z(\eta)\mathcal{P}\,
\mathbf m_{\rm SkN}(r,\varphi),
\label{eq:ASky_transformation}
\end{equation}
where $\mathcal{P}$ is an improper transformation with $\det \mathcal{P}=-1$. In the
conventional classification of topological spin textures in terms of
the topological charge $Q$, vorticity $n$, and helicity $\eta$
\cite{Zhang2020}, $\mathcal{P}$ reverses the winding of the in-plane
magnetization, changing the vorticity from $n=+1$ to $n=-1$ and
consequently reversing the sign of the topological charge. The
subsequent rotation $\mathcal{R}_z(\eta)$ generates configurations with different
helicities $\eta$, while leaving the vorticity and topological charge
unchanged. The exchange interaction and perpendicular magnetic
anisotropy remain unchanged under these transformations.
More generally, the canonical form of the resulting magnetic texture
depends on the order in which the spin-space rotations and the improper
transformation are applied. We therefore adopt the transformation
sequence consistent with the conventional classification of skyrmions
and antiskyrmions \cite{Zhang2020}. In particular, $\mathcal{P}$ does not commute
with the helicity rotation, $\mathcal{P} \mathcal{R}_z(\eta)=\mathcal{R}_z(-\eta)\mathcal{P}$. Consequently, the
order of the transformations determines how the helicity parameter
$\eta$ is assigned to the resulting magnetic configuration and thereby
fixes the canonical N\'eel- and Bloch-type configurations used
throughout this work.

The DMI must be transformed consistently with this improper spin-space
operation. In the DMI-tensor representation, the action of $\mathcal{P}$ involves
both the transformation of the spin indices and an additional
determinant factor associated with the transformation of the Lifshitz
invariants. The subsequent rotation $\mathcal{R}_z(\eta)$ then generates the
continuous helicity dependence. The resulting generalized DMI
combinations are summarized in Table~\ref{tab:transformation_rules},
while their complete derivation is provided in Supplementary Information S2.

\textit{Antibimerons.}
The antibimeron family combines the transformations introduced above,
\begin{equation}
\mathbf m_{\rm ABim}(r,\varphi,\eta)
=
\mathcal{R}_y\!\left(\frac{\pi}{2}\right)
\mathcal{R}_z(\eta)\mathcal{P}\,
\mathbf m_{\rm SkN}(r,\varphi).
\label{eq:AB_operator}
\end{equation}
Thus, the transformation sequence first generates the antiskyrmion
through $\mathcal{P}$, the rotation $\mathcal{R}_z(\eta)$ controls its helicity, and the
subsequent rotation $\mathcal{R}_y(\pi/2)$ maps the resulting texture onto the
corresponding antibimeron, rotating the easy axis from
$\hat{\mathbf z}$ to $\hat{\mathbf x}$. The exchange interaction
remains invariant, whereas the anisotropy and DMI are transformed
consistently with the spin configuration.

The resulting antibimeron DMI is given by the corresponding combination
of Lifshitz invariants listed in Table~\ref{tab:transformation_rules}.
The complete transformation of the magnetization, anisotropy, and DMI
is derived in Supplementary Information S2.

Together, these transformations establish a unified construction of
skyrmions, bimerons, antiskyrmions, and antibimerons from a single
reference configuration. The corresponding DMI energy densities can be
obtained consistently both by transforming the magnetization profile and
by independently transforming the Lifshitz-invariant representation.
Importantly, this framework is not tied to the particular analytical
ansatz used here and can be applied more generally to arbitrary
metastable magnetic textures, provided that the corresponding
micromagnetic energy functional is transformed consistently. The
transformation rules and the resulting generalized DMI energy densities
are summarized in Table~\ref{tab:transformation_rules}.

\section{Current-driven dynamics}

The spin-space transformation framework introduced in the previous
section naturally extends to the current-driven dynamics of the
generated magnetic textures. To establish the connection between the
proposed spin-space transformations and their transport properties, we
employ the collective-coordinate description based on the Thiele
equation \cite{Thiele_PRL_1973,Kamppeter_PRB_1999}. The complete
derivation of the generalized Thiele equation, the corresponding driving
force generated by the spin--orbit torque, and its analytical solution
are presented in Supplementary Information S4. Here, we summarize the
key results relevant to the transformation properties and
current-driven dynamics of the magnetic textures.

Unless stated otherwise, all micromagnetic simulations presented in this
work employ the material parameters reported in Refs.~\cite{Tretiakov_PhysRevB_2019,Hlushchenko2026PRB}: exchange stiffness
$A=15~{\rm pJ/m}$,
saturation magnetization
$M_s=0.58~{\rm MA/m}$,
DMI energy
scale
$D_{\rm DMI}=3~{\rm mJ/m^2}$,
effective uniaxial anisotropy constant
$|K_u|=0.8~{\rm MJ/m^3}$,
Gilbert damping
$\alpha=0.3$,
spin polarization
$P=0.35$,
ferromagnetic layer thickness
$\Delta_t=1~{\rm nm}$,
spin-polarization angle
$\psi=\pi/2$,
corresponding to current flowing along the $+x$ direction, and current
density
$j=4\times10^{11}~{\rm A/m^2}$.
Additional details of the micromagnetic implementation, including the
spin--orbit torque formulation and the generalized DMI fields, are
provided in Supplementary Information S3.

\begin{figure*}
\centering
\includegraphics[width=\linewidth]{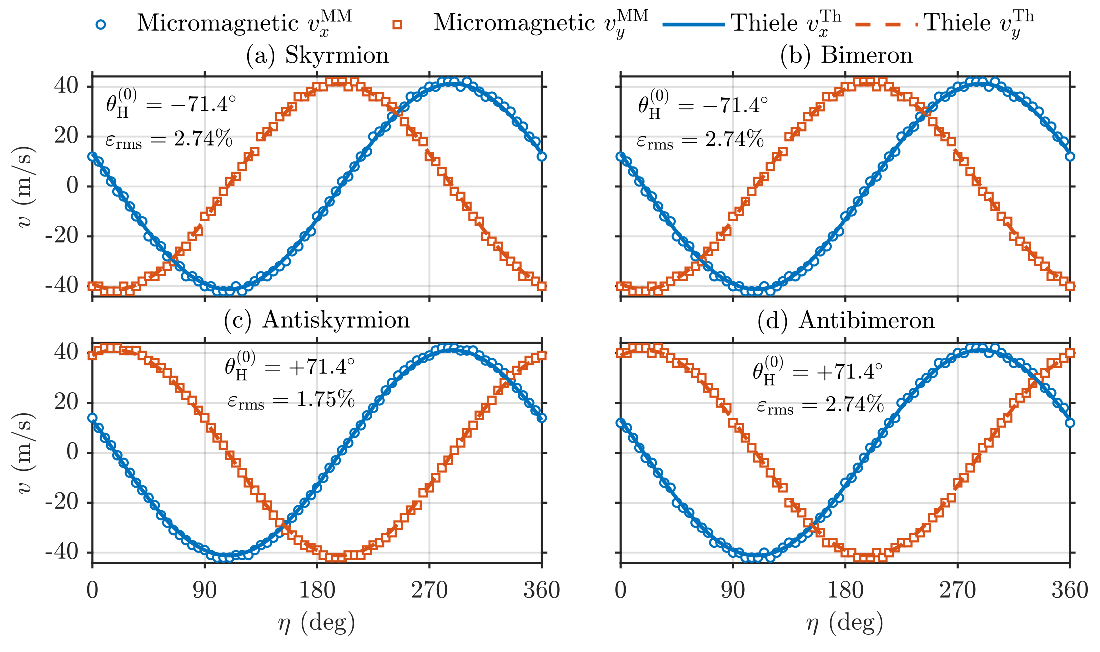}
\caption{\label{fig:current_driven_dynamics}
Dependence of the longitudinal ($v_x$) and transverse ($v_y$)
velocity components on the helicity angle $\eta$ for
(a) hybrid skyrmions,
(b) hybrid bimerons,
(c) hybrid antiskyrmions, and
(d) hybrid antibimerons.
Symbols represent the micromagnetic simulation results, while solid
and dashed lines show the corresponding predictions of the generalized
Thiele equation for $v_x$ and $v_y$, respectively.
The theoretical velocity magnitude is $V_{\rm Th}=41.56~{\rm m/s}$,
while the gyrotropic Hall angle of the canonical texture is
$\theta_{\rm H}^{(0)}=-71.4^\circ$ for skyrmions and bimerons and
$\theta_{\rm H}^{(0)}=+71.4^\circ$ for antiskyrmions and antibimerons.
The relative root-mean-square deviation between the micromagnetic
and theoretical velocities, $\varepsilon_{\rm rms}$, is indicated
for each texture.
The dynamics is driven by the spin--orbit torque using a spin
polarization of $P=0.35$, spin-polarization angle $\psi=\pi/2$,
corresponding to current flowing along the $+x$ direction, and current
density $j=4\times10^{11}~{\rm A/m^2}$.
}
\end{figure*}

Within the rigid-body approximation, the magnetic texture retains its
internal structure during its motion, and the LLG equation, including
the spin--orbit torque, reduces to
\begin{equation}
\mathbf G\times\mathbf v
+
\alpha
\hat{\mathbf D}\,
\mathbf v
=
\mathbf F,
\label{eq:Thiele}
\end{equation}
where
$\mathbf v=(v_x,v_y)$
is the propagation velocity,
$\mathbf G=(0,0,G)$
is the gyrovector,
$\hat{\mathbf D}$
is the dissipative tensor,
and
$\mathbf F$
is the generalized driving force generated by the spin--orbit torque.

Within the proposed spin-space transformation framework, the
current-driven dynamics is determined by the transformation properties
of the gyrovector, dissipative tensor, and generalized driving force.
The spin-space transformations impose simple relations between these
quantities under the elementary operations introduced in the previous
section. In particular, the dissipative tensor is unchanged by the
transformations considered here, whereas the operator $\mathcal P$
reverses the sign of the gyrovector. Thus,
\begin{equation}
\begin{aligned}
\hat{\mathbf D}^{\rm Sky}
&=
\hat{\mathbf D}^{\rm Bim}
=
\hat{\mathbf D}^{\rm ASky}
=
\hat{\mathbf D}^{\rm ABim},
\\
G^{\rm Sky}
&=
G^{\rm Bim}
=
-G^{\rm ASky}
=
-G^{\rm ABim}.
\end{aligned}
\label{eq:DG_transform}
\end{equation}
These transformation properties, together with the corresponding
transformation of the generalized driving force, directly lead to
universal velocity relations for the complete families of skyrmions,
bimerons, antiskyrmions, and antibimerons.

The micromagnetic calculations further show that the dissipative tensor
is nearly isotropic over the considered helicity range,
$D_{xx}\simeq D_{yy}=D$ and $D_{xy}\simeq0$. Consequently, the canonical
Hall angle is determined by the ratio of the gyrotropic and dissipative
contributions \cite{Mertig_PhysRevB_2019},
\begin{equation}
\theta_{\rm H}^{(0)}
=
-\arctan\!\left(\frac{G}{\alpha D}\right),
\label{eq:Hall_angle_0}
\end{equation}
where the gyrovector and dissipative tensor are evaluated from the
corresponding magnetization profiles, as described in Supplementary
Information S4. The helicity rotation then gives
\begin{equation}
\theta_{\rm H}(\eta)
=
\theta_{\rm H}^{(0)}-Q\eta,
\label{eq:Hall_angle_main}
\end{equation}
with $Q=\pm1$ denoting the topological charge. Using the corresponding
values of the gyrovector and dissipative tensor yields
$\theta_{\rm H}^{(0)}=-71.4^\circ$ for skyrmions and bimerons, and
$\theta_{\rm H}^{(0)}=+71.4^\circ$ for antiskyrmions and antibimerons.
The evaluation of these dynamical quantities and the analytical
derivation of the helicity-dependent Hall angle are given in
Supplementary Information S4.

To test these analytical predictions, we continuously varied the
helicity and directly compared the longitudinal ($v_x$) and transverse
($v_y$) velocity components obtained from micromagnetic simulations with
the corresponding solutions of the generalized Thiele equation. The
analytical expressions for the generalized driving force and the
resulting velocity components are derived in Supplementary Information
S4. The results are shown in Fig.~\ref{fig:current_driven_dynamics}.
Symbols denote the micromagnetic results, while solid and dashed lines
show the Thiele predictions for $v_x$ and $v_y$, respectively.

The Thiele description reproduces both velocity components over the
full helicity range for all four texture families. To quantify the
agreement, we evaluate the root-mean-square deviation between the
micromagnetic and theoretical velocity components over all simulated
helicities and normalize it by the corresponding root-mean-square
value of the micromagnetic velocities. The resulting relative error
$\varepsilon_{\rm rms}$ remains below $3\%$ for all four texture
families, with the individual values indicated in
Fig.~\ref{fig:current_driven_dynamics}. The definition of
$\varepsilon_{\rm rms}$ and the procedure used to extract the
micromagnetic velocities are given in Supplementary Information S4.
The theoretical velocity magnitude is obtained directly from the
Thiele equation by evaluating the corresponding gyrovector,
dissipative tensor, and spin--orbit-torque force integrals. For the
parameters used here, this gives
$V_{\rm Th}=41.56~{\rm m/s}$. The close agreement confirms that the
spin-space transformation framework consistently captures the
current-driven dynamics of skyrmions, bimerons, antiskyrmions, and
antibimerons, including the reversal of the Hall response associated
with the sign change of the gyrovector.

\section{Conclusions and Outlook}

In this work, we have developed a unified spin-space transformation
framework that systematically generates complete families of skyrmions,
bimerons, antiskyrmions, and antibimerons within a single continuum
micromagnetic description. Starting from a reference N\'eel skyrmion,
successive spin-space transformations produce hybrid magnetic textures
with arbitrary helicity while consistently transforming the
corresponding magnetic anisotropy, generalized
DMI, effective magnetic fields, and
micromagnetic energy functionals. Consequently, the proposed formalism establishes a unified
theoretical framework in which proper spin rotations generate
continuous helicity families within each topological-charge sector,
while an improper spin-space transformation relates the sectors with
opposite topological charge.

The developed formalism also provides a unified analytical description
of the current-driven dynamics of hybrid magnetic textures. The
generalized Thiele theory identifies helicity as an independent
geometric degree of freedom that continuously controls the Hall angle
while leaving the drift velocity nearly unchanged. Consequently, the
Hall response is no longer an intrinsic material property but becomes a
programmable parameter, enabling both complete suppression of the Hall
effect and deterministic steering of magnetic textures toward arbitrary
in-plane directions under a fixed driving current. This establishes
helicity as a practical control knob for engineering the trajectories
of topological magnetic textures, opening new opportunities for
programmable routing of magnetic information carriers, efficient
current-driven transport without transverse losses, and reconfigurable
spintronic architectures. Since the proposed framework is independent
of the microscopic mechanism responsible for stabilizing a particular
helicity, it naturally applies to hybrid magnetic textures realized
through DMI engineering, competing
magnetic interactions, magnetic anisotropy, or external stimuli,
making it directly relevant to emerging experimental platforms.

Beyond individual magnetic textures, the proposed framework naturally
extends to coupled multilayer systems, including synthetic
antiferromagnets, where each magnetic layer can independently undergo
the spin-space transformations introduced in this work. Such a generalization generates a substantially richer class of coupled magnetic textures with independently controllable helicities, chiralities, and
topological charges, thereby providing extra degrees of freedom
for tailoring their static and dynamical properties. This
perspective extends the concept of spin-space transformations well beyond
individual topological textures~\cite{Tretiakov_PhysRevB_2019} to coupled
multilayer systems and naturally complements recent advances in the
stabilization and current-driven dynamics of synthetic
antiferromagnetic skyrmions
\cite{Legrand_NatMater_2020,Pham_Science_2024}, as well as emerging
studies of interacting synthetic antiferromagnetic skyrmions with
different chiralities~\cite{deSouzaSilva_PRL_2025}. More broadly, it elevates
the present spin-space transformation framework from a means of relating magnetic textures to a general
design principle for helicity-engineered topological magnetic matter and
facilitates the development of future spintronic devices based on
programmable control of topological magnetic states.

\section*{Supporting Information}

\textbf{Supplementary Information:} Analytical derivations, generalized
DMI transformations, micromagnetic implementation, collective-coordinate
theory, and descriptions of Supplementary Movies S1 and S2 (PDF).

\textbf{Supplementary Movie S1:} Continuous spin-space transformations
of hybrid topological magnetic textures (MP4).

\textbf{Supplementary Movie S2:} Helicity-controlled current-driven
motion of a hybrid bimeron (MP4).

\section*{Acknowledgments}
\label{sec:ack}
A.V.H. and M.I.B. acknowledge support from the National Research Foundation of Ukraine under Project No. 2025.07/0086, Excellent Science in Ukraine (2026--2028). A.V.C. acknowledges support from the BMFTR project 01DK24006 PLASMA-SPIN-ENERGY.  O.A.T. acknowledges the support from the Australian Research Council (Grant No. DP200101027) and the NCMAS grant. J.W.K. acknowledges the support from the National Science Centre, Poland, under Grant No. 2020/39/O/ST5/02110.

\clearpage
\onecolumn

\pagestyle{fancy}
\fancyhf{}
\fancyhead[L]{Supplementary Information}
\fancyhead[R]{\textit{Helicity-Controlled Hall Transport}}
\fancyfoot[C]{\thepage}
\setlength{\headheight}{15pt}
\renewcommand{\headrulewidth}{0.4pt}

\setcounter{section}{0}
\setcounter{subsection}{0}
\setcounter{equation}{0}
\setcounter{figure}{0}
\setcounter{table}{0}
\renewcommand{\thesection}{S\arabic{section}.}
\renewcommand{\thesubsection}{S\arabic{section}.\arabic{subsection}.}
\renewcommand{\theequation}{S\arabic{equation}}
\renewcommand{\thefigure}{S\arabic{figure}}
\renewcommand{\thetable}{S\arabic{table}.}

\setlength{\parindent}{0pt}
\setlength{\parskip}{0.45em}

\begin{center}

{\Large\bfseries Supplementary Information}\\[1.5em]

{\large\bfseries
Helicity-Controlled Hall Transport in Hybrid Topological Magnetic Textures
}\\[1.5em]

Anton V. Hlushchenko$^{1}$,
Julia Kharlan$^{2,3,4}$,
Mykhailo I. Bratchenko$^{1}$,
Aleksei~V.~Chechkin$^{1,5,6}$,
Jaros{\l}aw W. K{\l}os$^{2}$
Oleg A. Tretiakov$^7$,

\begin{flushleft}
$^{1}$National Science Center \emph{"Kharkiv Institute of Physics and Technology"},
61108 Kharkiv, Ukraine

$^{2}$ISQI, Faculty of Physics, Adam Mickiewicz University,
61-614 Pozna{\'n}, Poland

$^{3}$V.G. Baryakhtar Institute for Magnetism,
National Academy of Sciences of Ukraine, 03142 Kyiv, Ukraine

$^{4}$G. V. Kurdyumov Institute for Metal Physics of the NAS of Ukraine, 02000 Kyiv, Ukraine

$^{5}$Max Planck Institute of Microstructure Physics,
06120 Halle, Germany

$^{6}$Faculty of Pure and Applied Mathematics,\\
Wroc{\l}aw University of Science and Technology,
50-370 Wroc{\l}aw, Poland

$^{7}$School of Physics, The University of New South Wales, 2052 Sydney, Australia

\end{flushleft}

\end{center}

\section*{Supplementary Summary}

This Supplementary Information provides the analytical and computational
details supporting the unified spin-space transformation framework for
hybrid skyrmions, bimerons, antiskyrmions, and antibimerons. It presents
the reference N\'eel-skyrmion ansatz and derives the generalized
Dzyaloshinskii--Moriya interactions generated by the corresponding
spin-space transformations. The DMI transformations are independently
verified using the DMI-matrix representation in terms of Lifshitz
invariants. The micromagnetic implementation and the
collective-coordinate theory underlying the helicity-controlled
current-driven dynamics and generalized Hall-angle relations are also
described. Finally, the accompanying supplementary movies are
summarized.

\tableofcontents

\newpage
\section{Reference N\'eel Skyrmion Ansatz}
\label{sec:S1}

Throughout this work, the canonical N\'eel skyrmion with helicity
$\eta=0$ is adopted as the reference magnetic texture from which all
other topological magnetic textures are generated through the spin-space
transformations. Following the conventional
classification of magnetic skyrmions summarized in Refs.~\cite{Zhang2020},
the normalized magnetization field of the reference configuration is
written in the axisymmetric form
\begin{equation}
\mathbf{m}(r,\varphi)=
\begin{pmatrix}
\sin\theta(r)\cos\varphi\\
\sin\theta(r)\sin\varphi\\
\cos\theta(r)
\end{pmatrix},
\label{eq:ansatz_original}
\end{equation}
where
$\varphi=\operatorname{atan2}(y,x)$
is the azimuthal angle in the film plane.

The radial magnetization profile is adopted from the analytical model
proposed in Refs.~\cite{Rohart2013,LU2023171044},
\begin{equation}
\theta(r)=
2\arctan\!\left(e^{\frac{r-R}{\Delta}}\right)
+
2\arctan\!\left(e^{\frac{r+R}{\Delta}}\right)
-\pi,
\label{eq:theta_original}
\end{equation}
where the equilibrium skyrmion size and characteristic domain-wall
width are given by
\begin{equation}
R=
\pi D_{\rm DMI}
\sqrt{
\frac{A}
{16AK_u^2-\pi^2D_{\rm DMI}^2K_u}
},
\label{eq:skyrmion_R}
\end{equation}
and
\begin{equation}
\Delta=
\frac{\pi D_{\rm DMI}}{4K_u},
\label{eq:skyrmion_Delta}
\end{equation}
respectively. Here, $A$ is the exchange stiffness, $D_{\rm DMI}$ is the DMI energy
scale, and $K_u$ is the effective perpendicular magnetic anisotropy. The parameters $R$ and $\Delta$
are obtained from the analytical model of Refs.~\cite{Rohart2013,LU2023171044},
which provides an approximate equilibrium profile for an isolated
N\'eel skyrmion.

For the chosen parametrization,
\begin{equation}
\theta(0)=0,
\qquad
\theta(\infty)=\pi.
\end{equation}
Thus, the magnetization points along $+z$ at the skyrmion center and
approaches $-z$ far from the core. In the domain-wall region,
$\theta\simeq\pi/2$, and for $\eta=0$ the in-plane magnetization points
radially outward, corresponding to the N\'eel-type helicity convention
adopted throughout this work.

The topological charge is defined as
\begin{equation}
Q=
\frac{1}{4\pi}
\int
\mathbf{m}\cdot
\left(
\partial_x\mathbf{m}
\times
\partial_y\mathbf{m}
\right)
\,dx\,dy.
\label{eq:Q_definition}
\end{equation}
For the axisymmetric ansatz of Eq.~(\ref{eq:ansatz_original}), the
topological charge density reduces to
\begin{equation}
\mathbf{m}\cdot
\left(
\partial_x\mathbf{m}
\times
\partial_y\mathbf{m}
\right)
=
\frac{1}{r}
\sin\theta(r)
\frac{d\theta(r)}{dr}.
\end{equation}
Consequently, Eq.~(\ref{eq:Q_definition}) becomes
\begin{equation}
Q=
\frac{1}{2}
\int_0^\infty
\sin\theta(r)
\frac{d\theta(r)}{dr}\,dr
=
\frac{1}{2}
\left[
\cos\theta(0)-\cos\theta(\infty)
\right].
\label{eq:Q_axisymmetric}
\end{equation}
Using the boundary conditions above gives
$Q=+1$. Thus, the reference N\'eel skyrmion belongs to the positive
topological-charge sector used throughout this work.

For the construction of the other magnetic-texture families, the same
radial profile $\theta(r)$ is retained and only the spin-space
transformations introduced in the main text are applied. For the construction of the other magnetic-texture families, the same
radial profile $\theta(r)$ is retained in the initial transformed
configurations, which are subsequently relaxed numerically in the
micromagnetic simulations.

\section{Derivation of the generalized Dzyaloshinskii--Moriya interaction}
\label{app:DMI}

The Dzyaloshinskii--Moriya interaction (DMI) provides the chiral energy
contribution responsible for stabilizing topological magnetic textures.
The proposed spin-space transformation
formalism generates the complete family of magnetic textures by
transforming the magnetization together with the corresponding DMI
energy density. This section derives the generalized DMI for the
reference skyrmion family, from which the remaining interactions follow
through the spin-space transformations.

Within continuum micromagnetics, two canonical DMI invariants are
commonly distinguished~\cite{TopologyInMagnetism2018}. The interfacial
DMI stabilizes N\'eel skyrmions in ultrathin magnetic films with broken
inversion symmetry~\cite{Sampaio2013}, whereas the bulk DMI favors Bloch
skyrmions in noncentrosymmetric chiral
magnets~\cite{Nagaosa2013}.

In the present work, the interfacial DMI is adopted as the reference
interaction. The complete family of DMI energy densities is then
generated analytically by global spin-space rotations of the
magnetization while leaving the spatial coordinate system unchanged.

The reference interfacial DMI energy density
\cite{TopologyInMagnetism2018,Niu_NatCommun_2024} is
\begin{equation}
w_{\rm SkN}
=
D_{\rm DMI}
\left[
-m_z\nabla\cdot\mathbf m
+
(\mathbf m\cdot\nabla)m_z
\right],
\label{eq:wN_vector}
\end{equation}
or, equivalently,
\begin{equation}
w_{\rm SkN}
=
D_{\rm DMI}
\Big[
-m_z\partial_xm_x
+
m_x\partial_xm_z
-
m_z\partial_ym_y
+
m_y\partial_ym_z
\Big].
\label{eq:wN_cartesian}
\end{equation}
Here, $D_{\rm DMI}$ denotes the DMI energy scale corresponding to the
two-dimensional micromagnetic energy density used throughout this work.

The corresponding Bloch-type contribution can be written as
\begin{equation}
w_{\rm SkB}
=
-D_{\rm DMI}\,
\mathbf m\cdot(\nabla\times\mathbf m),
\label{eq:wB_vector}
\end{equation}
which, for a two-dimensional film, becomes
\begin{equation}
w_{\rm SkB}
=
D_{\rm DMI}
\Big[
-m_z\partial_xm_y
+
m_y\partial_xm_z
+
m_z\partial_ym_x
-
m_x\partial_ym_z
\Big].
\label{eq:wB_cartesian}
\end{equation}

The reference N\'eel skyrmion is rotated in spin space according to
\begin{equation}
\mathbf m'(r,\varphi,\eta)
=
\mathcal{R}_z(\eta)\,
\mathbf m(r,\varphi,0),
\end{equation}
where $\mathcal{R}_z(\eta)$ is the rotation matrix describing a rotation by the
helicity angle $\eta$ about the $z$ axis,
\begin{equation}
\mathcal{R}_z(\eta)
=
\begin{pmatrix}
\cos\eta&-\sin\eta&0\\
\sin\eta&\cos\eta&0\\
0&0&1
\end{pmatrix}.
\end{equation}

Since the rotation is global and spatially uniform, the spatial
derivatives transform according to the same rotation matrix,
\begin{align}
m_x'
&=
m_x\cos\eta-m_y\sin\eta,
\nonumber\\
\partial_i m_x'
&=
\partial_i m_x\cos\eta
-
\partial_i m_y\sin\eta,
\nonumber\\
m_y'
&=
m_x\sin\eta+m_y\cos\eta,
\nonumber\\
\partial_i m_y'
&=
\partial_i m_x\sin\eta
+
\partial_i m_y\cos\eta,
\nonumber\\
m_z'
&=
m_z,
\nonumber\\
\partial_i m_z'
&=
\partial_i m_z,
\end{align}
where $i=x,y$.

Applying the corresponding spin-space transformation to the reference
interfacial DMI energy density, Eq.~(\ref{eq:wN_vector}), and collecting
the terms proportional to $\cos\eta$ and $\sin\eta$ yields
\begin{equation}
w_{\rm DMI}^{\rm Sky}(\eta)
=
w_{\rm SkN}\cos\eta
+
w_{\rm SkB}\sin\eta.
\label{eq:w_general}
\end{equation}
Thus, the helicity angle $\eta$ parametrizes the continuous spin-space
rotation of the reference N\'eel skyrmion together with the
corresponding generalized DMI interaction.

The four canonical helicities recover the corresponding N\'eel- and
Bloch-type DMI energy densities,
\begin{align}
w_{\rm DMI}^{\rm Sky}(0)
=
w_{\rm SkN},\quad
w_{\rm DMI}^{\rm Sky}\!\left(\frac{\pi}{2}\right)
=
w_{\rm SkB},\quad
w_{\rm DMI}^{\rm Sky}(\pi)
=
-w_{\rm SkN}, \quad
w_{\rm DMI}^{\rm Sky}\!\left(\frac{3\pi}{2}\right)
=
-w_{\rm SkB}.
\end{align}

The corresponding effective DMI field follows from the variational
derivative of the DMI energy functional,
\begin{equation}
\mathbf H_{\rm DMI}
=
-\frac{1}{\mu_0M_s}
\frac{\delta E_{\rm DMI}}
{\delta\mathbf m},
\qquad
E_{\rm DMI}
=
\int
w_{\rm DMI}\,dV.
\label{eq:HDMI}
\end{equation}

Substituting Eq.~(\ref{eq:w_general}) gives
\begin{equation}
\mathbf H_{\rm DMI}^{\rm Sky}(\eta)
=
\frac{2D_{\rm DMI}}{\mu_0M_s}
\left[
\mathbf H_{\rm SkN}\cos\eta
+
\mathbf H_{\rm SkB}\sin\eta
\right],
\label{eq:HDMI_general}
\end{equation}
where
\begin{align}
\mathbf H_{\rm SkN}
=
\begin{pmatrix}
-\partial_xm_z\\
-\partial_ym_z\\
\partial_xm_x+\partial_ym_y
\end{pmatrix},
\quad
\mathbf H_{\rm SkB}
=
\begin{pmatrix}
-\partial_ym_z\\
\partial_xm_z\\
-\partial_xm_y+\partial_ym_x
\end{pmatrix}.
\end{align}

Equation~(\ref{eq:HDMI_general}) provides the reference generalized DMI
field for the skyrmion family. The corresponding DMI energy densities
and effective fields for bimerons, antiskyrmions, and antibimerons are
obtained by applying the respective spin-space transformations
consistently to the DMI functional. In particular, the
orientation-reversing transformation $\mathcal{P}$ requires the corresponding
transformation of the chiral DMI structure.

\subsection{DMI-matrix representation and hybrid skyrmions}
\label{app:DMI_matrix}

An alternative and independent representation of the generalized
DMI is provided by the DMI-matrix formalism of
Ref.~\cite{Niu_NatCommun_2024}. In this representation, the DMI is
described by a $3\times3$ matrix
\begin{equation}
\hat{\mathcal D}
=
\begin{pmatrix}
D_{11} & D_{12} & D_{13}\\
D_{21} & D_{22} & D_{23}\\
D_{31} & D_{32} & D_{33}
\end{pmatrix},
\label{eq:DMI_matrix_general}
\end{equation}
where the first index specifies the orientation of the DMI vector,
$x$, $y$, or $z$, while the second index specifies the direction of
the corresponding bond vector or, in the continuum representation,
the spatial derivative. The DMI energy density can be expressed in
terms of the Lifshitz invariants
\begin{equation}
L_{ij}^{(k)}
=
m_i\partial_k m_j-m_j\partial_k m_i,
\label{eq:Lifshitz_Niu}
\end{equation}
as
\begin{align}
w_{\rm DMI}
={}&
D_{11}L_{yz}^{(x)}
+D_{12}L_{yz}^{(y)}
+D_{13}L_{yz}^{(z)}
\nonumber\\
&+
D_{21}L_{zx}^{(x)}
+D_{22}L_{zx}^{(y)}
+D_{23}L_{zx}^{(z)}
\nonumber\\
&+
D_{31}L_{xy}^{(x)}
+D_{32}L_{xy}^{(y)}
+D_{33}L_{xy}^{(z)} .
\label{eq:DMI_matrix_energy}
\end{align}
For the two-dimensional textures considered here, the terms involving
$\partial_z$ can be omitted. This representation provides a convenient
way of tracking the transformation of the DMI under spin-space
transformations independently of any particular magnetization ansatz.

For the reference N\'eel skyrmion, the DMI energy density used in the
present work,
\begin{equation}
w_{\rm SkN}
=
D_{\rm DMI}
\Big[
-m_z\partial_xm_x
+
m_x\partial_xm_z
-
m_z\partial_ym_y
+
m_y\partial_ym_z
\Big],
\label{eq:wSkN_matrix}
\end{equation}
corresponds to the DMI matrix
\begin{equation}
\hat{\mathcal D}_{\rm SkN}
=
D_{\rm DMI}
\begin{pmatrix}
0 & 1 & 0\\
-1 & 0 & 0\\
0 & 0 & 0
\end{pmatrix}.
\label{eq:Dmatrix_SkN}
\end{equation}
Indeed, the only nonzero components are
$D_{12}=D_{\rm DMI}$ and $D_{21}=-D_{\rm DMI}$, yielding
\begin{equation}
w_{\rm SkN}
=
D_{\rm DMI}(L_{yz}^{(y)}
-
L_{zx}^{(x)}),
\end{equation}
which is identical to Eq.~(\ref{eq:wSkN_matrix}).

For a global proper rotation of the spin space, $\det \mathcal{R}=+1$, the first
index of the DMI matrix transforms together with the spin components,
whereas the spatial coordinates remain unchanged. The corresponding
DMI-matrix transformation is therefore
\begin{equation}
\hat{\mathcal D}'
=
\mathcal{R}\,\hat{\mathcal D}.
\label{eq:Dmatrix_rotation}
\end{equation}
For the helicity rotation
\begin{equation}
\mathcal{R}_z(\eta)
=
\begin{pmatrix}
\cos\eta&-\sin\eta&0\\
\sin\eta&\cos\eta&0\\
0&0&1
\end{pmatrix},
\end{equation}
the reference N\'eel DMI matrix becomes
\begin{equation}
\hat{\mathcal D}_{\rm Sky}(\eta)
=
\mathcal{R}_z(\eta)\hat{\mathcal D}_{\rm SkN},
\end{equation}
or explicitly
\begin{equation}
\hat{\mathcal D}_{\rm Sky}(\eta)
=
D_{\rm DMI}
\begin{pmatrix}
\sin\eta & \cos\eta & 0\\
-\cos\eta & \sin\eta & 0\\
0&0&0
\end{pmatrix}.
\label{eq:Dmatrix_hybrid_skyrmion}
\end{equation}

The corresponding continuum DMI energy density is therefore
\begin{align}
w_{\rm DMI}^{\rm Sky}(\eta)
={}
D_{\rm DMI}\cos\eta
\left[
-L_{zx}^{(x)}
+
L_{yz}^{(y)}
\right]
+
D_{\rm DMI}\sin\eta
\left[
L_{yz}^{(x)}
+
L_{zx}^{(y)}
\right].
\end{align}
Using the definitions of the canonical N\'eel- and Bloch-type DMI
densities, this expression can be rearranged as
\begin{equation}
w_{\rm DMI}^{\rm Sky}(\eta)
=
w_{\rm SkN}\cos\eta
+
w_{\rm SkB}\sin\eta,
\label{eq:DMI_matrix_reproduces_sky}
\end{equation}
which exactly reproduces the generalized DMI obtained above by directly
rotating the magnetization. Thus, the helicity-dependent generalized
DMI admits an equivalent representation as a continuous rotation of
the DMI matrix in spin space.

\subsection{Antiskyrmions in the DMI-matrix representation}
\label{app:DMI_matrix_antiskyrmions}

The antiskyrmion family is generated from the corresponding skyrmion
family by the orientation-reversing transformation operator $\mathcal{P}$
introduced in the main text. In spin space, this transformation reverses
the $y$ component of the magnetization while leaving the other
components unchanged,
\begin{equation}
\mathcal{P}:
\begin{pmatrix}
m_x\\
m_y\\
m_z
\end{pmatrix}
\longrightarrow
\begin{pmatrix}
m_x\\
-m_y\\
m_z
\end{pmatrix}.
\label{eq:P_spin_matrix}
\end{equation}
The corresponding matrix representation is
\begin{equation}
\mathcal{P}=
\begin{pmatrix}
1&0&0\\
0&-1&0\\
0&0&1
\end{pmatrix},
\qquad
\det(\mathcal{P})=-1.
\label{eq:P_matrix}
\end{equation}
Unlike the proper spin-space rotations, $\mathcal{P}$ is an improper orthogonal
transformation. It reverses the winding of the in-plane magnetization
and changes the sign of the topological charge while preserving the
radial profile of the texture.

The transformation of the DMI requires additional care because the
DMI energy is expressed in terms of Lifshitz invariants, which contain
the antisymmetric combination of spin components and their spatial
derivatives. Consequently, under an improper spin-space transformation,
the DMI vector index transforms as a pseudovector. While the spatial
coordinates remain unchanged, the DMI matrix therefore transforms as
\begin{equation}
\hat{\mathcal D}'
=
\det(\mathcal{P})\,\mathcal{P}\hat{\mathcal D}.
\label{eq:Dmatrix_P_transformation}
\end{equation}
For the present transformation, $\det(\mathcal{P})=-1$, and hence
\begin{equation}
\hat{\mathcal D}'
=
-\mathcal{P}\hat{\mathcal D}.
\end{equation}

Applying this rule to the reference N\'eel skyrmion DMI matrix gives
\begin{equation}
\hat{\mathcal D}_{\rm ASkN}
=
\det(\mathcal{P})\,
\mathcal{P}\hat{\mathcal D}_{\rm SkN}
=
D_{\rm DMI}
\begin{pmatrix}
0&-1&0\\
-1&0&0\\
0&0&0
\end{pmatrix}.
\label{eq:Dmatrix_ASkN}
\end{equation}
Thus, the N\'eel antiskyrmion is characterized by the symmetric
in-plane DMI components
\begin{equation}
D_{12}=D_{21}=-D_{\rm DMI}.
\end{equation}
The corresponding DMI energy density is
\begin{equation}
w_{\rm ASkN}
=
D_{\rm DMI}
\left[
-L_{yz}^{(y)}
-
L_{zx}^{(x)}
\right],
\end{equation}
or, explicitly,
\begin{equation}
w_{\rm ASkN}
=
D_{\rm DMI}
\Big[
-m_z(\partial_xm_x-\partial_ym_y)
+
m_x\partial_xm_z
-
m_y\partial_ym_z
\Big],
\end{equation}
which corresponds to the anisotropic DMI sector associated with
antiskyrmions in the DMI-matrix classification of
Ref.~\cite{Niu_NatCommun_2024}.

The complete antiskyrmion family is obtained by subsequently applying
the helicity rotation $\mathcal{R}_z(\eta)$. Since $\mathcal{R}_z(\eta)$ is a proper
rotation, its action on the DMI matrix is simply given by left
multiplication. The complete transformation is therefore
\begin{equation}
\hat{\mathcal D}_{\rm ASky}(\eta)
=
\mathcal{R}_z(\eta)\,
\det(\mathcal{P})\,\mathcal{P}\,
\hat{\mathcal D}_{\rm SkN}.
\label{eq:Dmatrix_unified_ASky}
\end{equation}
Explicitly, this gives
\begin{equation}
\hat{\mathcal D}_{\rm ASky}(\eta)
=
D_{\rm DMI}
\begin{pmatrix}
\sin\eta&-\cos\eta&0\\
-\cos\eta&-\sin\eta&0\\
0&0&0
\end{pmatrix}.
\label{eq:Dmatrix_hybrid_ASky}
\end{equation}

The corresponding DMI energy density is
\begin{align}
w_{\rm DMI}^{\rm ASky}(\eta)
={}
D_{\rm DMI}
\left[
-L_{zx}^{(x)}
-
L_{yz}^{(y)}
\right]\cos\eta+
D_{\rm DMI}
\left[L_{yz}^{(x)}
-L_{zx}^{(y)}
\right]\sin\eta.
\label{eq:w_ASky_Lifshitz}
\end{align}
Equivalently,
\begin{equation}
w_{\rm DMI}^{\rm ASky}(\eta)
=
w_{\rm ASkN}\cos\eta
+
w_{\rm ASkB}\sin\eta.
\label{eq:w_ASky_general}
\end{equation}

For the four canonical helicities, the generalized DMI reduces to
\begin{align}
w_{\rm DMI}^{\rm ASky}(0)
=
w_{\rm ASkN},~
w_{\rm DMI}^{\rm ASky}\left(\frac{\pi}{2}\right)
=
w_{\rm ASkB},~
w_{\rm DMI}^{\rm ASky}(\pi)
=
-w_{\rm ASkN},~
w_{\rm DMI}^{\rm ASky}\left(\frac{3\pi}{2}\right)
=
-w_{\rm ASkB}.
\end{align}

Thus, the DMI-matrix representation reproduces the complete
hybrid-antiskyrmion family generated by the spin-space transformation
formalism. The additional factor $\det(\mathcal{P})$ is essential for improper
transformations and ensures the correct transformation of the chiral
DMI structure.

\subsection{Bimerons in the DMI-matrix representation}
\label{app:DMI_matrix_bimerons}

The bimeron can be viewed as the in-plane counterpart of a skyrmion.
A convenient construction starts from the axisymmetric skyrmion profile
introduced in Ref.~\cite{Zhang2020}, which is adopted as the reference
texture in the present work. The connection between skyrmions and
bimerons can then be established through a global rotation of the
magnetization in spin space. In particular,
Ref.~\cite{Tretiakov_PhysRevB_2019} demonstrated that a $90^\circ$
rotation of all magnetic moments of a skyrmion about the in-plane $y$
axis transforms the texture into a bimeron, with the easy-axis
direction rotated from the out-of-plane to the in-plane direction.

In the present formalism, this transformation is represented by
\begin{equation}
\mathcal{R}_y\left(\frac{\pi}{2}\right)
=
\begin{pmatrix}
0&0&1\\
0&1&0\\
-1&0&0
\end{pmatrix}.
\label{eq:SI_Ry_bimeron}
\end{equation}
Since the spatial coordinate system is kept fixed, the rotation acts on
the spin index of the DMI matrix according to
\begin{equation}
\hat{\mathcal D}_{\rm Bim}
=
\mathcal{R}_y\left(\frac{\pi}{2}\right)
\hat{\mathcal D}_{\rm SkN}.
\label{eq:Dmatrix_Bim_rotation}
\end{equation}
Starting from the reference N\'eel-skyrmion matrix,
\begin{equation}
\hat{\mathcal D}_{\rm SkN}
=
D_{\rm DMI}
\begin{pmatrix}
0&1&0\\
-1&0&0\\
0&0&0
\end{pmatrix},
\end{equation}
one obtains
\begin{equation}
\hat{\mathcal D}_{\rm BimN}
=
D_{\rm DMI}
\begin{pmatrix}
0&0&0\\
-1&0&0\\
0&-1&0
\end{pmatrix}.
\label{eq:Dmatrix_BimN}
\end{equation}
Thus, the canonical N\'eel bimeron is represented by the two nonzero
components
\begin{equation}
D_{21}=D_{32}=-D_{\rm DMI}.
\label{eq:Dmatrix_BimN_components}
\end{equation}

Substitution of Eq.~(\ref{eq:Dmatrix_BimN}) into the general
DMI-matrix representation gives
\begin{equation}
w_{\rm BimN}
=
D_{\rm DMI}
\left[
-L_{zx}^{(x)}
-
L_{xy}^{(y)}
\right].
\label{eq:w_BimN_matrix}
\end{equation}

The complete helicity-dependent bimeron family is obtained by applying
the same rotation to the helicity-dependent skyrmion DMI matrix. In
accordance with the spin-space transformation
$\mathcal{R}_y(\pi/2)\mathcal{R}_z(\eta)$, this gives
\begin{equation}
\begin{aligned}
\hat{\mathcal D}_{\rm Bim}(\eta)
=
\mathcal{R}_y\left(\frac{\pi}{2}\right)
\mathcal{R}_z(\eta)
\hat{\mathcal D}_{\rm SkN}
=
D_{\rm DMI}
\begin{pmatrix}
0&0&0\\
-\cos\eta&\sin\eta&0\\
-\sin\eta&-\cos\eta&0
\end{pmatrix}.
\end{aligned}
\label{eq:Dmatrix_Bim_eta}
\end{equation}

Consequently, the N\'eel- and Bloch-type bimeron contributions are
\begin{align}
w_{\rm BimN}
=
D_{\rm DMI}
\left[
-L_{zx}^{(x)}
-
L_{xy}^{(y)}
\right], \quad w_{\rm BimB}
=
D_{\rm DMI}
\left[
L_{xy}^{(x)}
-L_{zx}^{(y)}
\right].
\end{align}
The generalized DMI is therefore
\begin{equation}
w_{\rm DMI}^{\rm Bim}(\eta)
=
w_{\rm BimN}\cos\eta
+
w_{\rm BimB}\sin\eta,
\end{equation}
which reproduces the generalized DMI obtained directly from the
spin-space transformation of the magnetization.

\subsection{Antibimerons in the DMI-matrix representation}
\label{app:DMI_matrix_antibimerons}

The antibimeron family is obtained by applying the spin-space rotation
$\mathcal{R}_y(\pi/2)$ to the hybrid-antiskyrmion family derived above. This
provides a direct connection between the antiskyrmion and antibimeron
families and avoids introducing an independent DMI matrix for the
antibimerons.

Since $\mathcal{R}_y(\pi/2)$ is a proper spin-space rotation, it acts on the
DMI matrix by left multiplication. The complete transformation sequence
therefore reads
\begin{equation}
\hat{\mathcal D}_{\rm ABim}(\eta)
=
\mathcal{R}_y\left(\frac{\pi}{2}\right)
\mathcal{R}_z(\eta)
\det(\mathcal{P})\,\mathcal{P}
\hat{\mathcal D}_{\rm SkN}.
\label{eq:Dmatrix_unified_ABim}
\end{equation}
Equivalently, using the hybrid-antiskyrmion DMI matrix obtained above,
\begin{equation}
\hat{\mathcal D}_{\rm ABim}(\eta)
=
\mathcal{R}_y\left(\frac{\pi}{2}\right)
\hat{\mathcal D}_{\rm ASky}(\eta).
\end{equation}

The resulting DMI matrix is
\begin{equation}
\hat{\mathcal D}_{\rm ABim}(\eta)
=
D_{\rm DMI}
\begin{pmatrix}
0&0&0\\
-\cos\eta&-\sin\eta&0\\
-\sin\eta&\cos\eta&0
\end{pmatrix}.
\label{eq:Dmatrix_hybrid_ABim}
\end{equation}
Substitution into the general DMI-matrix representation gives
\begin{align}
w_{\rm DMI}^{\rm ABim}(\eta)
={}
D_{\rm DMI}
\left[
-L_{zx}^{(x)}
+
L_{xy}^{(y)}
\right]\cos\eta
+
D_{\rm DMI}
\left[
-L_{zx}^{(y)}
-
L_{xy}^{(x)}
\right]\sin\eta.
\label{eq:w_ABim_Lifshitz}
\end{align}
Thus, the generalized antibimeron DMI can be written as
\begin{equation}
w_{\rm DMI}^{\rm ABim}(\eta)
=
w_{\rm ABimN}\cos\eta
+
w_{\rm ABimB}\sin\eta.
\end{equation}
The helicity parameter $\eta$ therefore continuously interpolates
between the canonical N\'eel- and Bloch-type antibimeron
configurations.

The complete transformation sequence can be summarized as
\begin{equation}
\hat{\mathcal D}_{\rm SkN}
\xrightarrow{\;\det(\mathcal{P})\mathcal{P}\;}
\hat{\mathcal D}_{\rm ASkN}
\xrightarrow{\;\mathcal{R}_z(\eta)\;}
\hat{\mathcal D}_{\rm ASky}(\eta)
\xrightarrow{\;\mathcal{R}_y(\pi/2)\;}
\hat{\mathcal D}_{\rm ABim}(\eta).
\label{eq:DMI_transformation_ABim}
\end{equation}

\section{ Micromagnetic implementation}

To verify the proposed spin-space transformation framework,
micromagnetic simulations were performed by solving the
Landau--Lifshitz--Gilbert (LLG) equation supplemented by the
spin--orbit torque (SOT)
\cite{Landau_PhysSowjet_1935,Gilbert_IEEE_2004,Sampaio2013},
\begin{equation}
\label{eq:LLG}
\frac{\partial\mathbf m}{\partial t}
=
-\gamma
\mathbf m\times\mathbf H_{\rm eff}
+\alpha
\mathbf m\times
\frac{\partial\mathbf m}{\partial t}
+\mathbf T_{\rm SOT},
\end{equation}
where
$\mathbf m=\mathbf M/M_s$
is the normalized magnetization,
$\gamma=2.21\times10^{5}~{\rm m/(A\,s)}$
is the gyromagnetic ratio in the magnetic-field
($\mathbf H$) convention,
$\alpha$
is the Gilbert damping parameter,
and
$M_s$
is the saturation magnetization.

The spin--orbit torque is described by
\begin{equation}
\label{eq:SOT}
\mathbf T_{\rm SOT}
=
-\gamma H_{\rm DL}
\mathbf m\times(\mathbf m\times\mathbf p)
-\gamma H_{\rm FL}
\mathbf m\times\mathbf p,
\end{equation}
where
\begin{equation}
H_{\rm DL}
=
\frac{\hbar Pj}
{2e\mu_0M_s\Delta_t},
\qquad
H_{\rm FL}
=
\beta H_{\rm DL},
\end{equation}
$P$
is the spin polarization,
$j$
is the applied current density,
$\Delta_t$
is the ferromagnetic layer thickness,
and
$\beta$
denotes the relative strength of the field-like torque.
In the present simulations,
$\beta=0.3$.
The spin-polarization direction is
\[
\mathbf p=(\cos\psi,\sin\psi,0).
\]
The factor $\mu_0$ in $H_{\rm DL}$ arises from the use of
$\gamma$ in the $\mathbf H$-field convention; equivalently, the same
torque can be written using the electronic gyromagnetic ratio and the
corresponding magnetic-induction convention.

The effective magnetic field entering
Eq.~(\ref{eq:LLG})
is decomposed into the exchange, anisotropy, and
DMI contributions,
\begin{equation}
\mathbf H_{\rm eff}
=
\mathbf H_{\rm ex}
+
\mathbf H_{\rm ani}
+
\mathbf H_{\rm DMI}.
\end{equation}

The exchange and anisotropy fields retain their conventional form,
\begin{equation}
\begin{aligned}
\mathbf H_{\rm ex}
&=
\frac{2A}{\mu_0M_s}
\nabla^2\mathbf m,
\\[2mm]
\mathbf H_{\rm ani}
&=
\frac{2K_u}{\mu_0M_s}
(\mathbf m\cdot\mathbf n)\mathbf n,
\end{aligned}
\end{equation}
where
$A$
is the exchange stiffness,
$K_u$
is the uniaxial anisotropy constant,
and
$\mathbf n$
denotes the easy-axis direction.
For skyrmions and antiskyrmions the easy axis is oriented along the film
normal,
$\mathbf n=\mathbf e_z$,
whereas for bimerons and antibimerons it is rotated into the film plane,
$\mathbf n=\mathbf e_x$,
consistently with the proposed spin-space transformations.

Unlike the exchange interaction, which is invariant under global
spin-space rotations, and the anisotropy, whose analytical form is
preserved after the corresponding rotation of the easy axis, the
DMI transforms into different generalized
energy densities depending on the magnetic texture.
The corresponding DMI effective field is evaluated from the variational
derivative of the appropriate DMI energy density,
\begin{equation}
\label{eq:HDMI_S3}
\mathbf H_{\rm DMI}
=
-\frac{1}{\mu_0M_s}
\left(
\frac{\partial w_{\rm DMI}}{\partial\mathbf m}
-
\nabla\cdot
\frac{\partial w_{\rm DMI}}
{\partial(\nabla\mathbf m)}
\right),
\end{equation}
where
$w_{\rm DMI}$
denotes the DMI energy density corresponding to the considered magnetic
texture.
The DMI energy densities and effective fields for bimerons,
antiskyrmions, and antibimerons are then obtained directly by applying
the spin-space transformations introduced in Sec.~II, resulting in a
single unified micromagnetic implementation for all four families of
topological magnetic textures.

For a magnetic film of thickness $\Delta_t$, the interfacial and bulk
DMI contributions can be expressed in the thickness-normalized
two-dimensional form,
\begin{equation}
w_{\rm DMI}
=
D_N L_N+D_B L_B,
\end{equation}
where $L_N$ and $L_B$ denote the corresponding interfacial and bulk
Lifshitz-invariant combinations. The effective coefficients $D_N$ and
$D_B$ have identical units of ${\rm J/m^2}$. If $D_s$ is the
interfacial DMI coefficient with units of ${\rm J/m}$ and $D_v$ is the
bulk DMI coefficient with units of ${\rm J/m^2}$, the effective
coefficients are
\begin{equation}
D_N=\frac{D_s}{\Delta_t},
\qquad
D_B=D_v.
\end{equation}
We define the DMI energy scale
\begin{equation}
D_{\rm DMI}
=
\sqrt{D_N^2+D_B^2},
\end{equation}
and parameterize the relative interfacial and bulk contributions as
\begin{equation}
D_N=D_{\rm DMI}\cos\eta,
\qquad
D_B=D_{\rm DMI}\sin\eta,
\end{equation}
so that
\begin{equation}
\eta
=
\operatorname{atan2}(D_B,D_N)
=
\operatorname{atan2}(D_v\Delta_t,D_s).
\end{equation}
Thus, the same helicity parameter $\eta$ that characterizes the
spin-space rotation also specifies the relative weights of the
interfacial and bulk DMI contributions in the covariantly transformed
model.

Micromagnetic simulations were performed using the Micromagnetics Module
of COMSOL Multiphysics
\cite{comsolMicromag,micromagModule}.
Throughout this work, the material parameters are specified using the
standard micromagnetic notation adopted by MuMax3
\cite{mumax3},
whereas the coefficients employed in the COMSOL implementation are
\begin{equation}
A'
=
\frac{2A}{\mu_0M_s},
\quad
K'
=
\frac{2K_u}{\mu_0M_s},
\quad
D_{\rm DMI}'
=
-\frac{2D_{\rm DMI}}{\mu_0M_s},
\end{equation}
where the primed quantities denote the coefficients entering the
weak-form implementation
\cite{Hlushchenko2026PRB}.

Unless stated otherwise, all simulations employ the material parameters
reported in~\cite{Tretiakov_PhysRevB_2019,Hlushchenko2026PRB}:
exchange stiffness
$A=15~{\rm pJ/m}$,
saturation magnetization
$M_s=0.58~{\rm MA/m}$,
DMI energy scale
$D_{\rm DMI}=3~{\rm mJ/m^2}$,
uniaxial anisotropy constant
$|K_u|=0.8~{\rm MJ/m^3}$,
Gilbert damping
$\alpha=0.3$,
spin polarization
$P=0.35$,
field-like torque parameter
$\beta=0.3$,
ferromagnetic layer thickness
$\Delta_t=1~{\rm nm}$,
spin-polarization angle
$\psi=\pi/2$,
corresponding to current flowing along the
$+x$
direction,
and current density
$j=4\times10^{11}~{\rm A/m^2}$.
The magnetization dynamics was simulated in a racetrack geometry using
a uniform finite-element discretization with a characteristic mesh size
of approximately
$1\times1\times1~{\rm nm}^3$.

\begin{figure*}
\centering
\includegraphics[width=\linewidth]{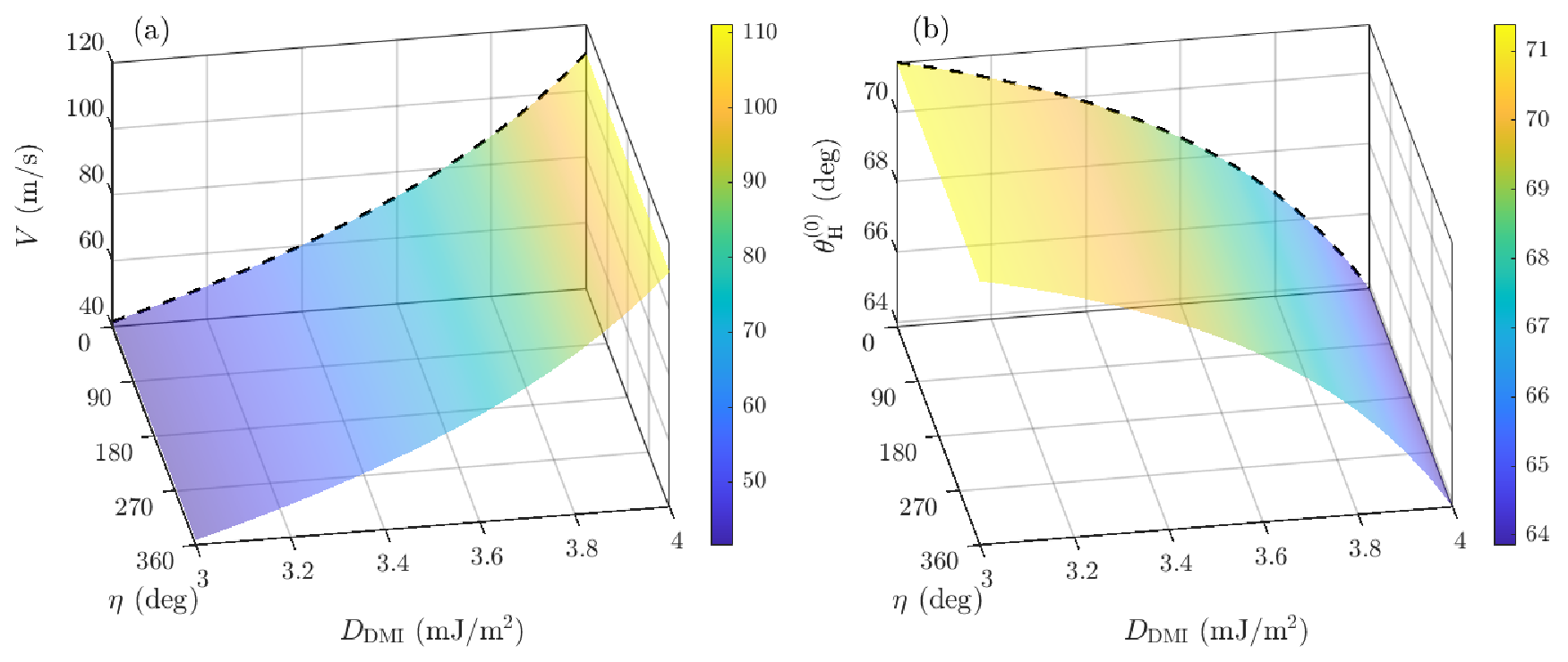}
\caption{\label{fig:helicity_dmi_dependence}
Dependence of (a) the velocity magnitude $V$ and (b) the canonical
gyrotropic Hall angle $\theta_{\rm H}^{(0)}$ on the helicity angle
$\eta$ and the DMI strength $D_{\rm DMI}$.
The results demonstrate that both quantities are independent of the
helicity angle $\eta$, while their values vary with the DMI strength.
The colored surfaces are obtained from the generalized Thiele equation
using the parameters extracted from micromagnetic simulations.
The dashed curves show the corresponding projections onto the
$D_{\rm DMI}$--$V$ and $D_{\rm DMI}$--$\theta_{\rm H}^{(0)}$ planes.
}
\end{figure*}

The present work focuses on the geometric role of helicity in controlling
Hall transport. We therefore keep the material parameters fixed when
comparing textures generated by the spin-space transformations, thereby
isolating the helicity-dependent contribution to the transport response.
The DMI energy scale is thus kept fixed in the main analysis. We note,
however, that the DMI strength can be varied independently of the
helicity. As illustrated in Fig.~\ref{fig:helicity_dmi_dependence},
systematic variation of both $\eta$ and $D_{\rm DMI}$ shows that the
velocity magnitude and the canonical gyrotropic Hall angle are
independent of $\eta$, whereas both quantities vary with the DMI
strength. Thus, the helicity dependence of the propagation direction
arises from the geometrical rotation of the driving force rather than
from a change in the intrinsic velocity magnitude or the gyrotropic
contribution to the Hall response.

The present calculations are formulated within a reduced local
micromagnetic model comprising exchange, uniaxial anisotropy, and
generalized DMI. Nonlocal dipolar interactions, finite-boundary effects,
and material-specific anisotropies may deform the transformed textures
and introduce quantitative deviations from the ideal covariance
relations. Such effects can be incorporated in a material-specific
realization without changing the underlying spin-space transformation
framework.

\section{Current-driven dynamics of hybrid magnetic textures}
\label{app:Thiele}

To describe the current-driven dynamics of the magnetic textures, we
employ the collective-coordinate approach based on the Thiele equation.
Within the rigid-body approximation, we assume that the magnetic texture
retains its internal profile while its center evolves in time.
Substituting this ansatz into the Landau--Lifshitz--Gilbert equation,
including the spin--orbit torque, and integrating over the magnetic
texture yields the Thiele equation
~\cite{Thiele_PRL_1973,Kamppeter_PRB_1999,Buttner_NatPhys_2015}.
Starting from this equation, we derive analytical expressions for the
current-driven dynamics of hybrid magnetic textures and, in particular,
for the generalized Hall angle introduced in the main text. The effects
of the proposed spin-space transformations enter through the generalized
driving force and the transformation properties of the gyrovector and
dissipative tensor.

\begin{figure*}
\centering
\includegraphics[width=\linewidth]{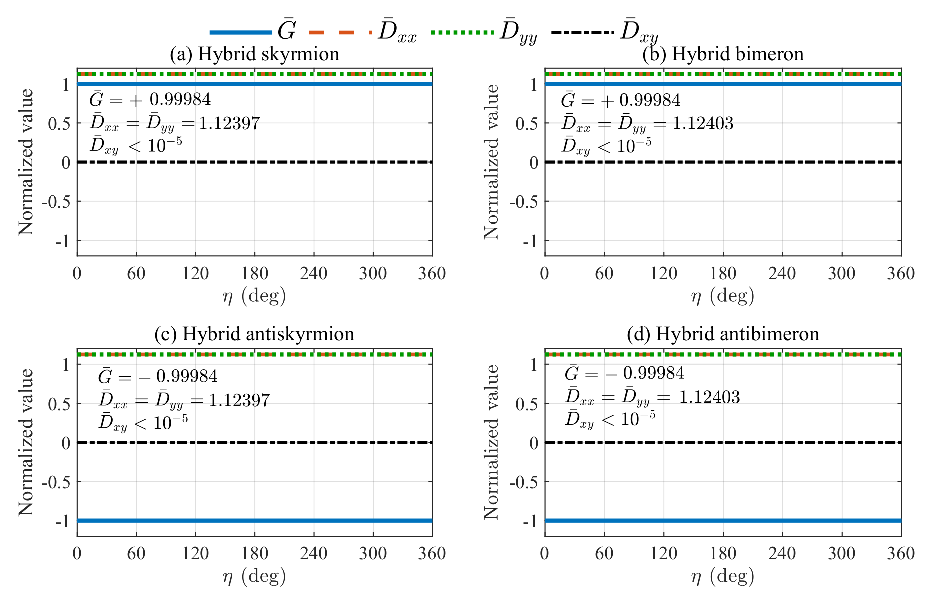}
\caption{\label{fig:gyro_dissipative_tensor}
Dependence of the normalized gyrovector $\bar G$ and dissipative tensor
components $\bar D_{xx}$, $\bar D_{yy}$, and $\bar D_{xy}$ on the
helicity angle $\eta$ for
(a) hybrid skyrmions,
(b) hybrid bimerons,
(c) hybrid antiskyrmions, and
(d) hybrid antibimerons.
The normalized quantities are defined as
$\bar G=\gamma G/(4\pi M_s\Delta_t)$ and
$\bar D_{ij}=\gamma D_{ij}/(4\pi M_s\Delta_t)$.
The gyrovector and dissipative tensor are calculated directly from the
relaxed micromagnetic configurations according to
Eqs.~(\ref{eq:G_appendix}) and
(\ref{eq:D_appendix}).
The simulations confirm that the gyrovector is independent of the
helicity and changes sign only upon reversal of the topological charge,
whereas the dissipative tensor remains practically invariant under the
proposed spin-space transformations, satisfying
$\bar D_{xx}\approx\bar D_{yy}\approx1.12$
and
$\bar D_{xy}\approx0$
throughout the entire helicity range.
}
\end{figure*}

\subsection{Thiele equation}

Within the rigid-body approximation, the magnetization is assumed to
preserve its internal structure during its motion,
\begin{equation}
\mathbf m(\mathbf r,t)
=
\mathbf m
\!\left(
\mathbf r-\mathbf R(t)
\right),
\end{equation}
where $\mathbf R=(X,Y)$ denotes the position of the texture center.

Substituting this ansatz into the LLG equation and integrating over the
magnetic texture yields the well-known Thiele equation
\begin{equation}
\mathbf G
\times
\mathbf v
+
\alpha
\hat{\mathbf D}\,
\mathbf v
=
\mathbf F,
\label{eq:Thiele_app}
\end{equation}
where $\mathbf v=(v_x,v_y)$ is the propagation velocity,
$\mathbf G=(0,0,G)$ is the gyrovector,
$\hat{\mathbf D}$ is the dissipative tensor,
and $\mathbf F$ is the generalized driving force generated by the
spin--orbit torque.

The gyrovector is defined as
\begin{equation}
\mathbf G
=
\frac{M_s\Delta_t}{\gamma}
\int
\mathbf m
\cdot
\left(
\partial_x\mathbf m
\times
\partial_y\mathbf m
\right)
\,dx\,dy\,
\mathbf e_z,
\label{eq:G_appendix}
\end{equation}
whereas the dissipative tensor is
\begin{equation}
D_{ij}
=
\frac{M_s\Delta_t}{\gamma}
\int
\partial_i\mathbf m
\cdot
\partial_j\mathbf m
\,dx\,dy.
\label{eq:D_appendix}
\end{equation}

The quantities entering
Eqs.~(\ref{eq:G_appendix}) and
(\ref{eq:D_appendix})
were evaluated numerically using the relaxed micromagnetic
configurations for the complete families of skyrmions, bimerons,
antiskyrmions, and antibimerons. The corresponding results are presented
in Fig.~\ref{fig:gyro_dissipative_tensor}. The gyrovector is found to be
independent of the helicity and changes only its sign upon reversal of
the topological charge, whereas the dissipative tensor remains nearly
isotropic throughout the entire helicity range, satisfying
$D_{xx}\approx D_{yy}$ and
$D_{xy}\approx0$
for all four texture families.

Consequently, the current-driven dynamics can be accurately described by
the isotropic approximation
$D_{xx}=D_{yy}\equiv D$
and
$D_{xy}=0$,
which forms the basis of the analytical derivation presented below.

The generalized force associated with the spin--orbit torque is obtained
by projecting the nonconservative torque onto the translational mode,
\begin{equation}
\label{eq:SOT_force_projection}
F_i^{\rm SOT}
=
\frac{\mu_0M_s\Delta_t}{\gamma}
\int
\left(
\mathbf m\times\partial_i\mathbf m
\right)
\cdot
\mathbf T_{\rm SOT}\,
dx\,dy .
\end{equation}
For the torque considered here, the damping-like contribution gives
\begin{equation}
\label{eq:SOT_force}
F_i
=
C_{\rm SOT}
I_{ij}p_j,
\qquad
C_{\rm SOT}
=
\frac{\hbar Pj}{2e},
\end{equation}
where the torque-response tensor is defined as
\begin{equation}
\label{eq:I_tensor}
I_{ij}
=
\int
\left[
\mathbf m\times\partial_i\mathbf m
\right]_j
\,dx\,dy .
\end{equation}
The field-like contribution proportional to $\beta$ gives a term
proportional to
$\int(\partial_i\mathbf m)\cdot\mathbf p\,dx\,dy$,
which is a boundary contribution. For the localized textures and
boundary conditions considered here, this contribution vanishes, so
that the translational force is determined by the damping-like torque. The gyrovector, dissipative tensor, and spin--orbit-torque force entering
the Thiele equation are evaluated directly from these integrals for the
relaxed micromagnetic configurations. Their values determine the
theoretical velocity magnitude without any fitting to the simulated
velocities.

The transformation properties of the driving force follow directly from
the transformation of the tensor $I_{ij}$. For a spin-space
transformation
$\mathbf m'=\mathcal O\mathbf m$,
with
$\mathcal O\in O(3)$,
one has
\begin{equation}
\mathbf m'\times\partial_i\mathbf m'
=
\det(\mathcal O)\,
\mathcal O
\left(
\mathbf m\times\partial_i\mathbf m
\right),
\end{equation}
and therefore
\begin{equation}
\label{eq:I_transformation}
I'
=
\det(\mathcal O)\,
I\mathcal O^{\mathsf T}.
\end{equation}
Equation~(\ref{eq:I_transformation}) shows explicitly how the spin-space
transformation modifies the direction of the SOT driving force while
preserving its magnitude in the ideal covariantly transformed model.
For the geometry considered here, $\mathbf p=\mathbf e_y$, so that
\begin{equation}
F_i=C_{\rm SOT}I_{iy}.
\end{equation}
For the reference N\'eel texture, the in-plane response tensor has the
form
\begin{equation}
I^{(0)}
=
I_0
\begin{pmatrix}
0 & 1\\
-1 & 0
\end{pmatrix},
\end{equation}
with $I_0>0$. Applying the corresponding spin-space transformations then
gives
\begin{equation}
\mathbf F_{\rm Sky,Bim}
=
F_0
\begin{pmatrix}
\cos\eta\\
-\sin\eta
\end{pmatrix},
\qquad
\phi_F^{\rm Sky,Bim}=-\eta,
\end{equation}
and
\begin{equation}
\mathbf F_{\rm ASky,ABim}
=
F_0
\begin{pmatrix}
\cos\eta\\
\sin\eta
\end{pmatrix},
\qquad
\phi_F^{\rm ASky,ABim}=+\eta,
\end{equation}
where
$F_0=C_{\rm SOT}I_0$.
Thus, the helicity dependence of the driving-force direction follows
directly from the transformation of the torque-response tensor.


\subsection{Generalized Hall effect of hybrid magnetic textures}

Using the isotropic form of the dissipative tensor established in the
previous subsection, the Thiele equation can be written as
\begin{equation}
\mathbf G
\times
\mathbf v
+
\alpha D\,\mathbf v
=
\mathbf F,
\label{eq:Thiele_reduced}
\end{equation}
where $\mathbf G=G\mathbf e_z$ is the gyrovector and $D$ is the
isotropic dissipative coefficient. To distinguish the direction of the
driving force from the spin-polarization direction, we parameterize the
force as
\begin{equation}
\mathbf F
=
F
(\cos\phi_F,\sin\phi_F),
\label{eq:F_appendix}
\end{equation}
where $F$ is the force magnitude and $\phi_F$ denotes its direction
with respect to the $+x$ axis.

Equation~(\ref{eq:Thiele_reduced}) gives
\begin{align}
v_x
&=
\frac{\alpha DF\cos\phi_F+GF\sin\phi_F}
     {(\alpha D)^2+G^2},
&
v_y
&=
\frac{\alpha DF\sin\phi_F-GF\cos\phi_F}
     {(\alpha D)^2+G^2}.
\label{eq:velocity_appendix}
\end{align}
The direction of the velocity, measured from the $+x$ axis, is therefore
given by
\begin{equation}
\phi_v
=
\operatorname{atan2}
\left(
v_y,v_x
\right)
=
\operatorname{atan2}
\left(
\alpha D\sin\phi_F-G\cos\phi_F,\,
\alpha D\cos\phi_F+G\sin\phi_F
\right),
\label{eq:velocity_angle}
\end{equation}
where $\operatorname{atan2}(y,x)$ retains the full angular range and
therefore uniquely determines the propagation direction.

Introducing the gyrotropic angle
\cite{Mertig_PhysRevB_2019}
\begin{equation}
\theta_0
=
\arctan
\left(
\frac{G}{\alpha D}
\right),
\label{eq:theta0}
\end{equation}
Eq.~(\ref{eq:velocity_angle}) can be written as
\begin{equation}
\phi_v
=
\phi_F-\theta_0
\pmod{2\pi}.
\label{eq:velocity_angle_general}
\end{equation}
Thus, the gyrotropic contribution rotates the propagation direction by
the angle $-\theta_0$ relative to the direction of the driving force.

For the current-driven geometry considered throughout this work, the
current flows along the $+x$ direction, while the spin polarization is
fixed at $\psi=\pi/2$, such that
$\mathbf p=\mathbf e_y$. For the reference N\'eel texture
($\eta=0$), the resulting SOT driving force is directed along the
current, $\phi_F=0$. Consequently, the canonical Hall angle, defined as
the propagation direction relative to the current direction, is
\begin{equation}
\theta_{\rm H}^{(0)}
=
-\theta_0
=
-\arctan
\left(
\frac{G}{\alpha D}
\right),
\label{eq:Hall0}
\end{equation}
in agreement with the conventional Thiele description of
current-driven skyrmion dynamics
\cite{Thiele_PRL_1973,Kamppeter_PRB_1999,Buttner_NatPhys_2015,Mertig_PhysRevB_2019}.

Within the proposed spin-space transformation framework, the helicity
changes the direction of the SOT driving force. For skyrmions and
bimerons, whose in-plane magnetization depends on the combination
$(\varphi+\eta)$, the force direction is rotated by $-\eta$ with
respect to the current,
\begin{equation}
\phi_F^{\rm Sky,Bim}=-\eta.
\label{eq:force_angle_SB}
\end{equation}
For antiskyrmions and antibimerons, characterized by the combination
$(\eta-\varphi)$, the corresponding force rotation has the opposite
sign,
\begin{equation}
\phi_F^{\rm ASky,ABim}=+\eta.
\label{eq:force_angle_ASAB}
\end{equation}
The propagation direction then follows directly from
Eq.~(\ref{eq:velocity_angle_general}), giving
\begin{equation}
\theta_{\rm H}^{\rm Sky,Bim}
=
-\arctan
\left(
\frac{G}{\alpha D}
\right)
-\eta
=
\theta_{\rm H}^{(0)}-\eta
\pmod{2\pi},
\label{eq:Hall_SB}
\end{equation}
for skyrmions and bimerons, and
\begin{equation}
\theta_{\rm H}^{\rm ASky,ABim}
=
-\arctan
\left(
\frac{G}{\alpha D}
\right)
+\eta
=
\theta_{\rm H}^{(0)}+\eta
\pmod{2\pi},
\label{eq:Hall_ASAB}
\end{equation}
for antiskyrmions and antibimerons.

These expressions reveal two distinct contributions to the Hall angle:
the gyrotropic term $-\theta_0$, determined by the ratio $G/(\alpha D)$,
and a purely geometrical contribution arising from the helicity-dependent
rotation of the driving force. The latter changes sign between the two
topological-charge sectors and provides continuous control of the
propagation direction through the helicity. In particular, the Hall
effect can be completely suppressed by choosing the helicity such that
the propagation direction is aligned with the current, corresponding to
$\theta_{\rm H}=0$ for forward propagation or $\theta_{\rm H}=\pi$ for
backward propagation.

The velocity magnitude is
\begin{equation}
v
=
\sqrt{v_x^2+v_y^2}
=
\frac{F}
{\sqrt{(\alpha D)^2+G^2}},
\label{eq:velocity_magnitude}
\end{equation}
and is independent of the direction of the driving force. Thus, within
the present fixed-parameter model, changing the helicity rotates the
propagation direction without changing the velocity magnitude.

Equations~(\ref{eq:Hall_SB}) and~(\ref{eq:Hall_ASAB}) provide a unified
collective-coordinate description of Hall transport for skyrmions,
bimerons, antiskyrmions, and antibimerons. Their agreement with the
micromagnetic simulations demonstrates that helicity acts as a geometric
control parameter for the current-driven transport of topological
magnetic textures.

\subsection{Velocity extraction and relative RMS deviation}

The micromagnetic velocity components were extracted from the simulated
trajectories of the texture center as
\begin{equation}
v_{x,i}^{\rm MM}
=
\frac{x(t_f)-x(t_0)}{t_f-t_0},
\qquad
v_{y,i}^{\rm MM}
=
\frac{y(t_f)-y(t_0)}{t_f-t_0},
\end{equation}
for each simulated helicity $\eta_i$.

The agreement between the micromagnetic results and the analytical
Thiele predictions was quantified using the root-mean-square deviation
over both velocity components and all simulated helicities,
\begin{equation}
{\rm RMSE}
=
\sqrt{
\frac{1}{N}
\sum_{i=1}^{N}
\left[
\left(
v_{x,i}^{\rm MM}
-
v_{x,i}^{\rm Th}
\right)^2
+
\left(
v_{y,i}^{\rm MM}
-
v_{y,i}^{\rm Th}
\right)^2
\right]
},
\end{equation}
where $N$ is the number of simulated helicities. To obtain a
dimensionless measure of the deviation, the RMSE was normalized by the
root-mean-square value of the micromagnetic velocity components,
\begin{equation}
\varepsilon_{\rm rms}
=
\frac{{\rm RMSE}}
{
\sqrt{
\frac{1}{N}
\sum_{i=1}^{N}
\left[
\left(v_{x,i}^{\rm MM}\right)^2
+
\left(v_{y,i}^{\rm MM}\right)^2
\right]
}
}.
\label{eq:relative_RMS_error}
\end{equation}
The quantity $\varepsilon_{\rm rms}$ provides a normalized measure of
the overall deviation between the micromagnetic results and the
corresponding analytical Thiele predictions.

\section{Supplementary Movie}

\subsection{Supplementary Movie S1: Continuous spin-space transformations}

Supplementary Movie~S1 illustrates the continuous evolution of hybrid
topological magnetic textures generated by spin-space transformations.
The movie presents the relaxed magnetic configurations obtained after
energy minimization in the absence of current for continuously varying
helicity
\[
0^\circ \le \eta < 360^\circ.
\]
The four panels correspond to

\begin{itemize}
\item[(a)] skyrmions,
\item[(b)] bimerons,
\item[(c)] antiskyrmions,
\item[(d)] antibimerons.
\end{itemize}

The movie demonstrates the continuous evolution from the canonical
N\'eel configurations through intermediate hybrid states to the
Bloch-type configurations and back to the initial state within each
topological-charge sector. It visualizes the unified spin-space
transformation framework, with proper spin rotations generating
continuous helicity families within a given topological sector, while
the sectors with opposite topological charge are related by an
improper spin-space transformation.

\subsection{Supplementary Movie S2: Helicity-controlled current-driven motion of a hybrid bimeron}

Supplementary Movie~S2 demonstrates the current-driven motion of a
hybrid bimeron for four characteristic helicities corresponding to
propagation along the four principal in-plane directions. The four
panels
show

\begin{itemize}
\item[(a)] propagation along the $+x$ direction for $\eta=289^\circ$,
\item[(b)] propagation along the $+y$ direction for $\eta=199^\circ$,
\item[(c)] propagation along the $-x$ direction for $\eta=109^\circ$,
\item[(d)] propagation along the $-y$ direction for $\eta=19^\circ$.
\end{itemize}

Each panel shows the current-driven dynamics over the time interval
from $t=0$ to $1~\mathrm{ns}$. The selected helicity angles follow from
the analytical solution of the generalized Thiele equation and correspond
to the four principal propagation directions predicted by the theory.

The movie illustrates that the propagation direction can be controlled
solely through the helicity parameter, while the propagation speed
remains essentially unchanged. The same helicity-controlled transport
mechanism applies to the other texture families generated by the
corresponding spin-space transformations.

\clearpage
\twocolumn
\bibliography{bimerons}
\end{document}